\documentclass[letterpaper, 10 pt, journal, twoside]{ieeetran}
\usepackage[utf8x]{inputenc}

\usepackage{graphicx, color}      
\usepackage{amsmath}
\usepackage{amssymb}
\usepackage[dvips]{epsfig}
\usepackage{amsthm}
\usepackage{enumerate}
\usepackage{mathabx}
\usepackage{tabularray}

\usepackage{tikz}

\newtheorem{remark}{Remark}
\newtheorem{definition}{Definition}

\newtheorem{proposition}{Proposition}

\newtheorem{problem}{Problem}

\newcommand{\der}{\frac{d}{dt}}
\newcommand{\derb}{\left(\frac{d}{dt}\right)}

\newcommand{\B}{\mathfrak{B}}
\newcommand{\R}{\mathbb{R}}

\DeclareMathOperator{\col}{col}

\newcommand{\N}{\mathbb{N}}

\newcommand{\mH}{\mathcal{H}}

\newcommand\copyrighttext{%
	\footnotesize \copyright 2024 IEEE. Personal use of this material is permitted. Permission from IEEE must be obtained for all other uses, in any current or future media, including reprinting/republishing this material for advertising or promotional purposes, creating new collective works, for resale or redistribution to servers or lists, or reuse of any copyrighted component of this work in other works.}
\newcommand\copyrightnotice{%
	\begin{tikzpicture}[remember picture,overlay]
		\node[anchor=south,yshift=7pt] at (current page.south) {\fbox{\parbox{\dimexpr\textwidth-\fboxsep-\fboxrule\relax}{\copyrighttext}}};
	\end{tikzpicture}%
}

\title{An input-output continuous-time version\\ of Willems’ lemma}
\author{Victor G. Lopez, \IEEEmembership{} Matthias A. M\"uller and \IEEEmembership{} Paolo Rapisarda  \IEEEmembership{}
\thanks{This work received funding from the European Research Council (ERC) under the European Union’s Horizon 2020 research and innovation programme (grant agreement No 948679).}
\thanks{V. G. Lopez and M. A. M\"uller are with the Leibniz University Hannover, Institute of Automatic Control, 30167 Hannover, Germany (e-mail: {\tt {lopez,mueller}@irt.uni-hannover.de}). }
\thanks{P. Rapisarda is with the School of Electronics and Computer Science, University of Southampton, Great Britain (e-mail: {\tt pr3@ecs.soton.ac.uk}).}}

\begin{document}
\maketitle
\thispagestyle{empty}
\pagestyle{empty}
\copyrightnotice

\begin{abstract}
We illustrate a novel version of Willems’ lemma for data-based representation of continuous-time systems. The main novelties compared to previous works are two. First, the proposed framework relies only on measured input-output trajectories from the system and no internal (state) information is required. Second, our system representation makes use of exact system trajectories, without resorting to orthogonal bases representations and consequent approximations. We first establish sufficient and necessary conditions for data-based generation of system trajectories in terms of suitable latent variables. Subsequently, we reformulate these conditions using measured input-output data and show how to span the full behavior of the system. Furthermore, we show how to use the developed framework to solve the data-based continuous-time simulation problem. 
\end{abstract}

\section{Introduction}
Parametrizing system trajectories is an essential  step of data-driven control methods, since in such a framework the design of controllers and control inputs can only be performed on the basis of such description of the plant dynamics. For linear, time-invariant, discrete-time systems such a parametrization was provided in \cite{WRM}, applied to \emph{data-driven simulation} in \cite{Markovsky08} and extended to multiple data sets in \cite{Waarde20}. Many different applications of this result for control design have been obtained, as described in the survey \cite{MarDor21}, and various extensions to more general (discrete-time) system classes have been developed such as, e.g., stochastic systems \cite{FauEtAl23}, linear parameter varying systems \cite{VerEtAl21} or classes of nonlinear systems \cite{RueSch20,AlsEtAl23}. 

More recently, two analogous parametrizations for continuous-time systems have been proposed respectively in \cite{LM} and \cite{REtAl23b}. Both approaches have been leveraged for control design, see \cite{LopMul24,REtAl23c}. In spite of their attractive theoretical properties and their applicability to control design, none of these results can be considered as the precise continuous-time counterpart of the fundamental lemma in \cite{WRM}. On the one hand, in \cite{LM} the conditions for generating system trajectories from data are given in terms of the internal state of the system (see \cite[Lemma 2]{LM}), which must be measured along with input-output trajectories. On the other hand, \cite{REtAl23b} uses orthogonal bases representations of the continuous-time trajectories, which implies that the resulting trajectories generated with this method are inherently an approximation of the true trajectories. It is desirable instead that  data-based continuous-time system trajectories can be  computed exactly and that they depend only on input-output measurements (here regarded as the \emph{external variables}). 

In this work we make some progress towards these goals. Combining some of the ideas of \cite{LM} and \cite{REtAl23a}, in this paper we provide the following contributions: 
\begin{itemize}
\item We illustrate a new perspective on some of the results in \cite{LM} by relating the differential equations that are required for spanning system trajectories to the concept of ``jets'' of input-output trajectories and their derivatives.
\item We extend the approach of \cite{LM} for exact data-based trajectory generation to the case of input-output measurements.
\item We sketch some of the connections between the approaches of \cite{LM}, \cite{REtAl23a} and \cite{REtAl23b}. 
\item We show how to use the proposed method to solve the data-based simulation problem.
\end{itemize}
The paper is organized as follows: in Section \ref{sec:probstat} we formulate the problem of trajectory-parametrization for continuous-time systems. In Section \ref{sec:mainresults} we define the concept of sufficient informativity for identification and, after 
establishing some instrumental results, we formulate a higher-order, input-output version of the fundamental lemma, generalizing the approach of \cite{LM}.  In Section \ref{sec:DDSim} we show how to use the obtained results to solve the data-driven simulation problem. In Section \ref{sec:concl} we summarize our findings and discuss some of the research directions currently being pursued. 

\subsection*{Notation}

We denote by $\R$ and $\N$ the sets of real and natural numbers, respectively. $\R[s]$ is the ring of polynomials with real coefficients, and $\R^{g\times q}[s]$ is the set of $g\times q$ matrices with entries in $\R[s]$. Polynomials and polynomial differential operators with constant coefficients are associated with each other: if $p_0+\ldots+p_L s^L\in\R[s]$, then we define $p\derb$ by $p\derb:=p_0+\ldots+p_L\frac{d^L}{dt^L}$. This notation extends in a natural way to polynomial matrices. 

$\R^n$ denotes the space of $n$-dimensional vectors with real entries.  $\R^{n\times m}$ denotes the set of $n\times m$ matrices with real entries. The transpose of a matrix $M$ is denoted by $M^\top$. If $M$ has full row rank, $M^\dagger$ denotes a right inverse. If $A$ and $B$ are two matrices with the same number of columns, we define $\col(A,B):=\begin{bmatrix}A^\top &B^\top \end{bmatrix}^\top$.

Given a continuous-time signal $w: \R \rightarrow \R^q$, fix $M\in\N$ and $T\in\R$, and denote by $\mathcal{H}(w)$ the time-dependent matrix (see also \cite{LM}) 
\begin{eqnarray}\label{eq:mHw}
\mathcal{H}(w)&:=&\begin{bmatrix} w(\cdot)& w(\cdot+T)&\ldots& w(\cdot+MT)\end{bmatrix}\nonumber\\
&=:&\begin{bmatrix} w(\cdot)& \sigma_T w(\cdot)&\ldots& \sigma_{MT}w(\cdot)\end{bmatrix}\; ,
\end{eqnarray}
where $\sigma$ represents the time-shift operator defined by 
\begin{eqnarray*}
    \sigma_\tau w(t):=w(t+\tau)\; , t\in\R\; .
\end{eqnarray*}
The $L$-th derivative of $w$ is denoted as $w^{(L)}:= \frac{d^L}{dt^L} w$.

\section{Problem statement}\label{sec:probstat}
In the behavioral framework, a dynamical system is described by its behavior $\B$, which consists of all input-output system trajectories  that are admissible by the system dynamics, see \cite{Willems1986}. In this paper, we consider a linear differential behavior $\B$ with $m$ inputs and $p$ outputs, i.e., $u \in \R^m$, $y \in \R^p$.

 In \cite[Theorem~2]{LM} it was shown that, under mild conditions, a data-based system representation of a controllable continuous-time system $\B$ is given by
 \begin{equation} \label{eq:pre1}
     \col \left(\overline u, \overline y \right) := \begin{bmatrix} \overline u \\ \overline y
 \end{bmatrix} = \begin{bmatrix} \mathcal{H}\left(u\right) \\ \mathcal{H}\left(y\right)
 \end{bmatrix} \alpha,
 \end{equation}
 where $\mathcal{H}(u)$, $\mH(y)$ are defined as in \eqref{eq:mHw},  $(u, y) \in \B$ are the data collected from the system, $(\overline u, \overline y) \in \B$ is any input-output system trajectory and $\alpha: \R \rightarrow \R^{M+1}$ satisfies the differential equation (see \cite[Lemma~2]{LM})
 \begin{equation} \label{eq:pre2}
 \begin{bmatrix} \mathcal{H}\left(u\right) \\ \mathcal{H}\left(x\right)
 \end{bmatrix} \alpha^{(1)} = - \begin{bmatrix} \mathcal{H}\left(u^{(1)}\right) \\ 0
 \end{bmatrix} \alpha + \begin{bmatrix} \overline u^{(1)} \\ 0
 \end{bmatrix}.
 \end{equation}
 Here, $x \in \R^n$ corresponds to a state variable for the system $\B$ that must be measurable for (\ref{eq:pre2}) to be solved in practice. Equations \eqref{eq:pre1}-\eqref{eq:pre2} imply that every possible input-output trajectory of the system $\B$ can be computed in a data-based fashion by performing suitable time-varying linear combinations using the collected input-state-output data.

In this paper we obtain alternative conditions that allow to represent the behavior $\B$ measuring only input-output data. That is, we wish to avoid the use of state information in our expression of $\B$. This is formalized in the following problem.

\begin{problem}
    Using only measurements of the external variables, determine sufficient and necessary conditions for the data-based system representation of continuous-time systems.
\end{problem}

In the following section, we present our solution to this problem. First, we obtain sufficient and necessary conditions to generate system trajectories in terms of a latent variable. Then, we use the obtained insight to state  sufficient and necessary conditions in terms of (measured) input-output data.

\section{Main results}\label{sec:mainresults}

\subsection{Data-based representation of controllable systems}


If the differential linear behavior $\B$ is controllable (see Definition 5.2.2, p. 153 of \cite{YellowBook}), then it can be described in \emph{image form} as 
\begin{equation}\label{eq:im}
\begin{bmatrix}
u\\y
\end{bmatrix}=\begin{bmatrix}
D_0+D_1 \der+\ldots+D_L \frac{d^L}{dt^L}\\ N_0+N_1 \der+\ldots+N_L \frac{d^L}{dt^L}
\end{bmatrix}\ell\; ,
\end{equation}
where $\ell$ is a $d$-dimensional latent variable, $L\in\N$, and $D_i$, $N_i$, $i=0,\ldots,L$, are matrices of appropriate dimensions. Without loss of generality, we  assume that $\ell$ is observable from $u$ and $y$. Notice that \eqref{eq:im} can be rewritten as
\begin{eqnarray}\label{eq:im2}
\begin{bmatrix}
u\\y
\end{bmatrix}&=&\begin{bmatrix}
D_0&D_1&\ldots&D_L\\ N_0&N_1&\ldots&N_L 
\end{bmatrix}
\begin{bmatrix}
\ell\\
\ell^{(1)}\\
\vdots\\
\ell^{(L)}
\end{bmatrix}\; .
\end{eqnarray}

For fixed $M\in\N$ and $T \in \R$, and using the signals $u$, $y$ and $\ell$, we define the time-varying matrices $\mH(u)$, $\mH(y)$ and $\mH(\ell)$ as in \eqref{eq:mHw}. Assuming that $u$ is sufficiently smooth, we also define analogous matrices for the $i$-th derivative of $u$, $y$ and $\ell$, denoted as $\mH(u^{(i)})$, $\mH(y^{(i)})$ and $\mH(\ell^{(i)})$, respectively.  Notice that the derivative of $\mH(u)$ is given by 
\begin{eqnarray}\label{eq:dermHu}
\der\left(\mathcal{H}(u)\right)&:=&\begin{bmatrix} \der \left(u\right)(\cdot)& \ldots& \der\left(\sigma_{MT}u\right)(\cdot)\end{bmatrix}\; .
\end{eqnarray}
Since differentiation and time-shift are commutative, we conclude that 
\[
\der\left(\mathcal{H}(u)\right)=\mathcal{H}\left( u^{(1)}\right)\; .
\]
Analogous definitions and equalities hold for the matrices  $\mathcal{H}(y)$ and $\mathcal{H}(\ell)$. 

From these  definitions and \eqref{eq:im2} we conclude that 
\[
\begin{bmatrix}
    \mathcal{H}(u) \\ \mathcal{H}(y)
\end{bmatrix}=\begin{bmatrix}
D_0&D_1&\ldots&D_L\\ N_0&N_1&\ldots&N_L 
\end{bmatrix} \begin{bmatrix}
\mathcal{H}(\ell)\\
\mathcal{H}(\ell^{(1)})\\
\vdots\\
\mathcal{H}(\ell^{(L)})
\end{bmatrix}\; .
\]

Now, let $\alpha:\R\rightarrow \R^{M+1}$ and define the new signals
\[
\overline{u}:=\mathcal{H}(u)\alpha \mbox{\rm~ and } \overline{y}:=\mathcal{H}(y) \alpha\; . 
\]

From the derivations above we conclude that 
\begin{equation}\label{eq:Hua}
\overline{u}=\mathcal{H}(u)\alpha=\begin{bmatrix}
D_0&D_1&\ldots&D_L\end{bmatrix} \begin{bmatrix}
\mathcal{H}(\ell)\\
\mathcal{H}(\ell^{(1)})\\
\vdots\\
\mathcal{H}(\ell^{(L)})
\end{bmatrix}\alpha\; ,
\end{equation}
and 
\begin{equation}\label{eq:Hya}
\overline{y}=\mathcal{H}(y)\alpha=\begin{bmatrix}
N_0&N_1&\ldots&N_L 
\end{bmatrix}
 \begin{bmatrix}
\mathcal{H}(\ell)\\
\mathcal{H}(\ell^{(1)})\\
\vdots\\
\mathcal{H}(\ell^{(L)})
\end{bmatrix}\alpha\; . 
\end{equation}

We now state necessary and sufficient conditions for $\col(\overline{u},\overline{y})$ defined by \eqref{eq:Hua}-\eqref{eq:Hya} to  belong to $\B$. These conditions are given in terms of the latent variable $\ell$. 
\begin{proposition}\label{prop:diffprop}
Let $\col(\overline{u},\overline{y})$ be defined by \eqref{eq:Hua}-\eqref{eq:Hya} with $\alpha:\R\rightarrow \R^{M+1}$. Assume that the latent variable $\ell$ is observable from the external variables $u$, $y$.  

The following statements are equivalent: 
\begin{enumerate}
\item  $\col(\overline{u},\overline{y})\in\B$; 
\item $\der\left( \mathcal{H}(\ell^{(i)}) \alpha\right)=\mathcal{H}(\ell^{(i+1)}) \alpha$, $i=0,\ldots,L-1$;
\item $\mathcal{H}(\ell^{(i)}) \alpha^{(1)}=0$, $i=0,\ldots,L-1$. 
\end{enumerate}
\end{proposition}
\begin{proof}
We prove first the equivalence of statements $1)$ and $2)$. Since $\col(\overline u, \overline y)$ is given by \eqref{eq:Hua}-\eqref{eq:Hya}, it can be written as
\begin{equation} \label{eq:proof1}
\col(\overline{u},\overline{y})=\begin{bmatrix} 
D_0&\ldots&D_L \\N_0&\ldots&N_L 
\end{bmatrix}\begin{bmatrix}
\mathcal{H}(\ell)\\
\mathcal{H}(\ell^{(1)})\\
\vdots\\
\mathcal{H}(\ell^{(L)})
\end{bmatrix}\alpha\; .
\end{equation}
Moreover, from \eqref{eq:im} we know that $\col(\overline{u},\overline{y})\in \B$ if and only if there exists a latent variable trajectory $\ell^\prime$ such that 
\begin{equation} \label{eq:proof2}
\col(\overline{u},\overline{y})=\begin{bmatrix}
D_0+D_1 \der+\ldots+D_L \frac{d^L}{dt^L}\\ N_0+N_1 \der+\ldots+N_L \frac{d^L}{dt^L}
\end{bmatrix}\ell^\prime\; .
\end{equation}
Subtracting \eqref{eq:proof1} from \eqref{eq:proof2} and applying the observability assumption, we conclude that 
\[
\mathcal{H}(\ell^{(i)}) \alpha=\ell^{\prime(i)}\; , \; i=0,\ldots,L\; . 
\]
This implies 
\begin{eqnarray*}
\der\left( \mathcal{H}(\ell^{(i)}) \alpha\right)&=&\der \left(\ell^{\prime(i)}\right)=\ell^{\prime (i+1)}\\&=&\mathcal{H}(\ell^{(i+1)}) \alpha\; , \; i=0,\ldots,L-1\; .
\end{eqnarray*}
This concludes the proof of $1) \Longleftrightarrow 2)$. 

To prove the equivalence of $2)$ and $3)$,  apply Leibniz's rule 
\[
\der\left( \mathcal{H}(\ell^{(i)}) \alpha\right)=\mathcal{H}(\ell^{(i+1)}) \alpha+\mathcal{H}(\ell^{(i)}) \alpha^{(1)}\; ;
\]
conclude that $2)$ holds if and only if $\mathcal{H}(\ell^{(i)}) \alpha^{(1)}=0$. 
\end{proof}


In Proposition \ref{prop:diffprop}  a characterization of system trajectories is given: the trajectories $\bar{u},\, \bar{y}$ computed via \eqref{eq:Hua}-\eqref{eq:Hya} are an admissible input-output trajectory for the system if and only if $\alpha$ satisfies any of the conditions in statements 2) or 3). Such conditions  cannot be verified directly from the input-output data, since the latent variable $\ell$ corresponding to a given input-output trajectory is in general not available for measurement. Moreover, since no assumption is made on the data $u,\, y$ used in \eqref{eq:Hua}-\eqref{eq:Hya} (e.g., informativity), Proposition~\ref{prop:diffprop} does not provide a data-driven description of the  behavior $\B$. Consequently the result is of limited use when considering the data-driven simulation problem, that consists of finding an output trajectory corresponding to a given input and initial conditions on the external variables.  

In order to state a data-based system characterization in terms of the external variables only, we need first to define the notion of sufficiency of information. 

\subsection{Sufficiently informative external trajectories}
The importance of the ``jets" of system trajectories in the characterization of linear differential systems has been suggested in \cite{Willems1986} and has been elegantly established in  
 \cite{Lomadze12} for the case of infinitely differentiable functions and of solution spaces ``with sufficiently many smooth functions" (for the definition, see Remark 2, p. 817, in \cite{Lomadze12}). 
 
 In \cite{REtAl23a,REtAl23b} jets have been applied to characterize the concept of ``persistency of excitation" and to provide a version of Willems' lemma based on representations of system trajectories in the space of the  sequences of their Chebyshev coefficients. It has been shown that a natural framework in which to pose any question related to continuous-time data (persistency of excitation, identifiability, ``sufficient informativity", etc.) {is that consisting of finite ``jets" generated by system trajectories $\col(u,y)$}: 
\begin{equation}\label{eq:jet}
J_L(u,y):=\begin{bmatrix} u^\top&u^{(1)\top}&\ldots &u^{(L)\top}&y^\top&\ldots&y^{(L)\top}\end{bmatrix}^\top \; ,
\end{equation}
where one  assumes that $L\geq \ell ag(\B)$, the lag of the system. Thus, the concept of \emph{jet} refers to the organization of the input-output trajectories and its derivatives as the vector-valued function of time in \eqref{eq:jet}. In the following we denote by $J_L(\B)$ the set of \emph{admissible $L$-jets}: 
\[
J_L(\B):=\left\{J_L(u,y) \mid \col(u,y)\in\B \right\}\; .
\]
Each  jet \eqref{eq:jet}, together with the corresponding shifts and time intervals defined as in  \eqref{eq:mHw}, defines a \emph{data matrix}: 
\begin{equation}\label{eq:datamatrix}
\begin{bmatrix}
\mH(u)\\ \mH(u^{(1)})\\ \vdots \\ \mH(u^{(L)})\\ \mH(y)\\ \mH(y^{(1)})\\ \vdots \\ \mH(y^{(L)})
\end{bmatrix}\; .
\end{equation}

\begin{remark}\rm
	Note that \eqref{eq:datamatrix} requires knowledge of the derivatives of the input and output trajectories of up to the $L$-th order. This is a common requirement in the continuous-time system identification literature; methods to compute an approximation of these derivatives have been investigated (see \cite{UnbRao98} and the discussion in \cite[Section V]{REtAl23a}). In the remainder of this paper, we assume availability of these derivative signals.
\end{remark}

\begin{definition}\label{def:SI}
Let $\eta_i\in\R^{1\times m}$, $\theta_i\in\R^{1\times p}$, $i=0,\ldots,L$, for some $L\geq \ell ag(\B)$. Moreover, consider a trajectory $\col(u,y)\in\B$ and fix the values $M \in \N$, $T\in \R$. The tuple $(\col(u,y),M,T)$ is \emph{sufficiently informative for identification} if, for each fixed $t^\prime$,  the equality 
\[
\begin{bmatrix}\eta_0&\ldots&\eta_L&\theta_0&\ldots&\theta_L \end{bmatrix}\begin{bmatrix}
	\mH(u(t^\prime))\\ \vdots \\ \mH(u^{(L)}(t^\prime))\\ \mH(y(t^\prime))\\ \vdots \\ \mH(y^{(L)}(t^\prime))
\end{bmatrix}=0 \;
\]
implies 
\[
\sum_{i=0}^L \eta_i \overline{u}^{(i)}(t) +\sum_{i=0}^L \theta_i \overline{y}^{(i)}(t)=0 \; ,
\]
for all $\col(\overline{u},\overline{y})\in\B$ and all $t$. 
\end{definition}
Definition \ref{def:SI} states that the data $\col(u,y)$ organized in a matrix as in \eqref{eq:datamatrix} is sufficiently informative for identification if, for each fixed time $t'$, the left annihilators of the matrix are in one-one correspondence with the differential equations that describe $\B$. Since $\col(u,y)\in\B$,  the set of left-annihilators of \eqref{eq:datamatrix} for \emph{any} selection of $M \in \N$ and $T\in \R$ \emph{contains} the set of annihilators of $\B$. Data informativity holds if the \emph{converse inclusion} holds, i.e. if the matrix \eqref{eq:datamatrix}  contains enough information to uniquely determine the generating behavior.

Note that the condition in Definition \ref{def:SI} can be verified from data by checking whether the matrix \eqref{eq:datamatrix} has the same rank at all times. It can be shown (similarly as in Proposition~2 of \cite{REtAl23a}) that this rank must be equal to $m(L+1)+n$, where $n$ is the dimension of a minimal internal state of $\B$.

\begin{remark}\rm 
    Establishing sufficient conditions for informativity for identification is often achieved by using \emph{persistently  excitating input signals}, see \cite{REtAl23a,LM} for alternative definitions. In this paper, we aim to derive sufficient and necessary conditions for data-based system representation of continuous-time systems without the need of using persistently exciting inputs. This is similar in spirit to the  relaxations of the conditions of Willems' lemma developed in \cite{MarDor23} for the discrete-time case. Designing persistently exciting inputs to guarantee the conditions in Definition \ref{def:SI} will be investigated elsewhere.
\end{remark}

\subsection{From input-output data to admissible $L$-jets}
The following result is a data-based characterization of linear differential systems analogous to that provided in \cite{LM}, based only on the external variables $u$ and $y$.

\begin{proposition}\label{prop:diffpropker}
Let $\col(u,y)\in\B$ and define the data matrix \eqref{eq:datamatrix}. Let $\alpha:\R\rightarrow \R^{M+1}$ be continuously differentiable. Define an $(m+p)(L+1)$-dimensional function by   
\begin{equation}\label{eq:alphajet}
\begin{bmatrix}
\mH(u)\\ \vdots \\ \mH(u^{(L)})\\ \mH(y)\\ \vdots \\ \mH(y^{(L)})
\end{bmatrix}\alpha \; .
\end{equation}
The following statements are equivalent: 
\begin{enumerate}
\item  \eqref{eq:alphajet} is an admissible $L$-jet; 
\item The following equations hold for $i=0,\ldots,L-1$: 
\begin{eqnarray}\label{eq:kercond}
\der\left( \mathcal{H}(u^{(i)}) \alpha\right)&=&\mathcal{H}(u^{(i+1)}) \alpha\nonumber\\
\der\left( \mathcal{H}(y^{(i)}) \alpha\right)&=&\mathcal{H}(y^{(i+1)}) \alpha\; ;
\end{eqnarray}
\item The following equations hold for $i=0,\ldots,L-1$: 
\begin{eqnarray}\label{eq:kercond2}
\mathcal{H}(u^{(i)}) \alpha^{(1)}&=&0\nonumber\\
\mathcal{H}(y^{(i)}) \alpha^{(1)}&=&0\; .
\end{eqnarray}
\end{enumerate}
Moreover, if $(\col(u,y),M,T)$ is sufficiently informative for identification, then any $\overline{u},\, \overline{y}$ with $\overline{u}$ continuously differentiable satisfies $\col(\overline{u},\overline{y}) \in \B$ if and only if its $L$-jet $J_L(\overline{u},\overline{y})$ can be written as in \eqref{eq:alphajet} with $\alpha \in  \R^{M+1}$ satisfying the statements in 2) and 3).
\end{proposition}
\begin{proof}
The equivalence of $2)$ and $3)$ follows from the equalities for $i=0,\ldots,L-1$: 
\begin{eqnarray*}
\der\left(\mH(u^{(i)})\alpha\right)&=&\mH(u^{(i+1)})\alpha+\mH(u^{(i)})\alpha^{(1)}\\
\der\left(\mH(y^{(i)})\alpha\right)&=&\mH(y^{(i+1)})\alpha+\mH(y^{(i)})\alpha^{(1)}\;.
\end{eqnarray*}

We prove $1) \Longrightarrow 2)$. Since \eqref{eq:alphajet} is an admissible $L$-jet,  there exists $\col(u^\prime,y^\prime)\in\B$ such that $u^{\prime (i)}=\mH(u^{(i)})\alpha$ and $y^{\prime (i)}=\mH(y^{(i)})\alpha$, $i=0,\ldots,L$. It follows that for $i=0,\ldots,L-1$ 
\begin{eqnarray*}
u^{\prime (i+1)}=\mH(u^{(i+1)})\alpha&=&\der \left(u^{\prime (i)}\right)=\der\left(\mH(u^{(i)})\alpha\right)\\
y^{\prime (i+1)}=\mH(y^{(i+1)})\alpha&=&\der \left(y^{\prime (i)}\right)=\der\left(\mH(y^{(i)})\alpha\right)\;.
\end{eqnarray*}

To prove that $2) \Longrightarrow 1)$, observe that the $(m+p)(L+1)$-dimensional function defined by 
\[
\begin{bmatrix}
f_0\\
\vdots\\
f_L\\
g_0\\
\vdots\\
g_L
\end{bmatrix}:=\begin{bmatrix}
\mH(u)\\ \vdots \\ \mH(u^{(L)})\\ \mH(y)\\ \vdots \\ \mH(y^{(L)})
\end{bmatrix}\alpha \; ,
\]
is the $L$-jet of $\col(f_0,g_0)$ since $2)$ implies that $f_i=f_0^{(i)}$ and $g_i=g_0^{(i)}$, $i=0,\ldots,L-1$. That such jet is admissible follows from the fact that the set of left-annihilators of \eqref{eq:datamatrix} contains the set of annihilators of $\B$ and, consequently, $\col(f_0,g_0)$ satisfies the differential equations describing $\B$.

We prove the last statement. Consider any trajectories $\overline{u}, \, \overline{y}$. From the implication $2) \Longrightarrow 1)$, if $J_L(\overline{u},\overline{y})$ can be written as in \eqref{eq:alphajet} with $\alpha$ satisfying $2)$ and $3)$, then $\col(\overline{u},\overline{y}) \in \B$. Conversely, consider any admissible $L$-jet $J_L(\overline{u},\overline{y})$ of the system with continuously differentiable $\overline{u}$. Since the data $u,y$ is sufficiently informative for identification, the set of left annihilators of \eqref{eq:datamatrix} at each time $t^\prime$ is isomorphic with the set of annihilators of $\B$. This implies that $J_L(\overline{u},\overline{y})$ can be written as \eqref{eq:alphajet} for some continuously differentiable $\alpha \in \R^{M+1}$. The proof is completed with the fact $1) \Longrightarrow 2)$ shown above.
\end{proof}

\begin{remark}\rm
Proposition \ref{prop:diffpropker} allows to generate every $(\overline{u},\overline{y})\in\B$ with continuously differentiable  $\overline{u}$. A similar differentiability requirement was made in \cite{LM}, which was later relaxed to include piecewise continuously differentiable inputs \cite[Corollary 1]{LM}. Such relaxation was  possible because of the  availability of state measurements assumed in that work, and will be investigated elsewhere for our input-output setting. 
\end{remark}

\begin{remark}\rm
In Proposition \ref{prop:diffpropker} it is not necessary to assume controllability of $\B$ as required in \cite{LM,REtAl23b}, since we  assume that the data are sufficiently informative for identification; controllability is needed if the data-based system representation is obtained using a persistently exciting input.
\end{remark}

The result in Proposition \ref{prop:diffpropker} provides a data-based characterization of continuous-time system trajectories alternative to those in \cite{LM,REtAl23b}. The main difference with the results in \cite{REtAl23b} is that we do not  use orthogonal bases representations. When using orthogonal bases in practice, one must truncate the series representation; hence, such method inherently provides only \emph{approximate} expressions for the system trajectories. Instead, the mathematical representation of system trajectories by means of \eqref{eq:alphajet} is exact, as is that obtained in \cite{LM}. The most important difference between the results of \cite{LM} and Proposition~\ref{prop:diffpropker} is that the latter does not require measuring the state as required in \cite[Lemma~2]{LM}. However, it is important to highlight that Proposition \ref{prop:diffpropker} is not an input-output version of the results in \cite{LM}. Other important differences distinguish the two methods, as we analyze in the following subsection.

\subsection{Relations with the results in \cite{LM}}
The developments in \cite{LM} make use of input-state representations of continuous-time systems
\begin{equation}\label{eq:behstate}
\B:=\left \{\col(u,x) \mid \der x=Ax+Bu \right\}\; ,
\end{equation}
and it is assumed that the state $x \in \R^n$ is directly measurable. 

Under these assumptions, choosing $L=\ell ag(\B)=1$, the jet \eqref{eq:jet} associated with $\col(u,x)\in\B$ is 
\[
J_1(u,x)=\begin{bmatrix}
u^\top & u^{(1)\top} & x^\top & x^{(1)\top}
\end{bmatrix}^\top\; .
\]
From Proposition \ref{prop:diffpropker}, we derive the following result. 
\begin{proposition}\label{prop:diffpropkerstate}
Let $\col(u,x)$ be a trajectory of \eqref{eq:behstate}; define the data matrix by 
\begin{equation}\label{eq:dataLMjet}
\begin{bmatrix}
\mH(u)\\
\mH\left(u^{(1)}\right)\\
\mH(x)\\
\mH\left(x^{(1)}\right)
\end{bmatrix}\; . 
\end{equation}
Let $\alpha:\R\rightarrow \R^{M+1}$ be continuously differentiable and define a $2(m+n)$-dimensional function by   
\begin{equation}\label{eq:alphajetstate}
\begin{bmatrix}
\mH(u)\\
\mH\left(u^{(1)}\right)\\
\mH(x)\\
\mH\left(x^{(1)}\right)
\end{bmatrix}\alpha \; .
\end{equation}
The following statements are equivalent: 
\begin{enumerate}
\item  \eqref{eq:alphajetstate} is an admissible $1$-jet for \eqref{eq:behstate}; 
\item The following equations hold: 
\begin{equation*}
\der\left( \mathcal{H}(u) \alpha\right)=\mathcal{H}(u^{(1)}) \alpha \mbox{\rm~and~}
\der\left( \mathcal{H}(x) \alpha\right)=\mathcal{H}(x^{(1)}) \alpha\; ;
\end{equation*}
\item The following equations hold: 
\begin{equation*}
\mathcal{H}(u) \alpha^{(1)}=0 \mbox{\rm~and~} \mathcal{H}(x) \alpha^{(1)}=0\; .
\end{equation*}
\end{enumerate}
\end{proposition}

The condition $\mathcal{H}(u) \alpha^{(1)}=0$ appearing in statement  $3)$ of Proposition \ref{prop:diffpropkerstate} does not appear in the analogous result Lemma 2 of \cite{LM}. This occurs since in our setting we work with {jets} consisting of the derivatives of $x$ \emph{and} $u$, while in \cite{LM} trajectories are generated from the matrix $\begin{bmatrix}
\mH(u)\\
\mH(x)\\
\mH\left(x^{(1)}\right)
\end{bmatrix}$, where data associated with the derivative of $u$ are \emph{not} present. The additional condition $\mathcal{H}(u) \alpha^{(1)}=0$ appearing in Proposition~\ref{prop:diffpropkerstate} is required since in our setting we need to impose that the derivative of $\mathcal{H}(u) \alpha$ equals $\mathcal{H}(u^{(1)}) \alpha$.

Note that the results in \cite{LM} can only be used when the internal state $x$ is available, i.e., if $y=x$. However, when this is the case the procedure in \cite{LM} can be applied without the need to compute the derivative of $x$ (compare Section \ref{sec:probstat}). On the other hand, Proposition \ref{prop:diffpropker} can be applied regardless of the availability of the state, although trajectory derivatives must be approximated.

\section{The data-based simulation problem} \label{sec:DDSim}
The data-driven simulation problem consists of finding an output trajectory corresponding to given initial conditions on the external variables trajectories and a given input trajectory. 

Using Proposition \ref{prop:diffpropker}, this problem is solved as follows.

\begin{proposition}\label{prop:simulationprob}
	Let $\col(u,y)\in\B$ and assume that $(\col(u,y),M,T)$ is sufficiently informative for identification. Consider an input trajectory $\bar u$ at least $L+1$ times differentiable with $L \geq \ell ag(\B)$. Moreover, suppose that the output initial conditions $\bar y^{(i)}(0)$, $i=0,\ldots,L$, are available. Then, the output trajectory $\overline{y}$ corresponding to these initial conditions and the input $\bar u$ is given by $\bar y(t) = \mH(y(t)) \alpha(t)$, where $\alpha:\R\rightarrow \R^{M+1}$ is a solution to the differential equation 
	\begin{equation} \label{eq:difeq2}
		\begin{bmatrix}
			\mathcal{H}(u) \\ \vdots \\ \mathcal{H}(u^{(L-1)}) \\ \mathcal{H}(u^{(L)}) \\ \mathcal{H}(y) \\ \vdots \\ \mathcal{H}(y^{(L-1)})
		\end{bmatrix} \alpha^{(1)} = -\begin{bmatrix}
			0 \\ \vdots \\ 0 \\ \mathcal{H}(u^{(L+1)}) \\ 0 \\ \vdots \\ 0
		\end{bmatrix} \alpha + \begin{bmatrix}
			0 \\ \vdots \\ 0 \\ \bar{u}^{(L+1)} \\ 0 \\ \vdots \\ 0
		\end{bmatrix}
	\end{equation}
	with initial conditions given by
	\begin{equation} \label{eq:initcond}
		\begin{bmatrix}
			\mathcal{H}(u(0)) \\ \vdots \\ \mathcal{H}(u^{(L)}(0)) \\ \mathcal{H}(y(0)) \\ \vdots \\ \mathcal{H}(y^{(L)}(0))
		\end{bmatrix} \alpha(0) = \begin{bmatrix}
			\bar u(0) \\ \vdots \\ \bar u^{(L)}(0) \\ \bar y(0) \\ \vdots \\ \bar y^{(L)}(0)
		\end{bmatrix}\; .
	\end{equation}
\end{proposition}
\begin{proof}
From Proposition \ref{prop:diffpropker}, we know that there exists an $\alpha$ such that the jet $\begin{bmatrix} \bar u^\top&\ldots &\bar u^{(L)\top}& \bar y^\top&\ldots& \bar y^{(L)\top}\end{bmatrix}^\top$ is given by \eqref{eq:alphajet}. Moreover, from \eqref{eq:kercond2} we have that \eqref{eq:alphajet} is an admissible $L$-jet for $\B$ if and only if
\begin{equation} \label{eq:difeq1}
	\begin{bmatrix}
		\mathcal{H}(u) \\ \vdots \\ \mathcal{H}(u^{(L-1)}) \\ \mathcal{H}(y) \\ \vdots \\ \mathcal{H}(y^{(L-1)})
	\end{bmatrix} \alpha^{(1)} = \begin{bmatrix}
	0 \\ \vdots \\ 0 \\ 0 \\ \vdots \\ 0
\end{bmatrix}\: .
\end{equation}
Notice that \eqref{eq:difeq1} does not contain information about the input $\bar u$. This is solved by noticing from \eqref{eq:alphajet} that $\bar u^{(L)} = \mH(u^{(L)}) \alpha$. Taking the derivative on both sides of this equality, we get
\begin{equation*}
	\bar u^{(L+1)} = \mathcal{H}(u^{(L)}) \alpha^{(1)} + \mathcal{H}(u^{(L+1)}) \alpha\; .
\end{equation*} 
Including this set of equations in \eqref{eq:difeq1}, we obtain \eqref{eq:difeq2}.
\end{proof}

In \eqref{eq:difeq2} we have a system of \emph{implicit} differential equations which can be solved using standard software packages for the numerical solution of differential equations. 

In some cases the differential equations \eqref{eq:difeq2} can be reformulated and solved more easily (see also Section IV-C in \cite{LM}). Assume that the data matrix on the left-hand side of \eqref{eq:difeq2} has full row rank (this assumption is discussed below); in such case, it is easy to see that solving the differential equation
\begin{multline} \label{eq:simulation}
	\alpha^{(1)} = -\begin{bmatrix}
		\mathcal{H}(u) \\ \vdots \\ \mathcal{H}(u^{(L-1)}) \\ \mathcal{H}(u^{(L)}) \\ \mathcal{H}(y) \\ \vdots \\ \mathcal{H}(y^{(L-1)})
	\end{bmatrix}^\dagger \begin{bmatrix}
		0 \\ \vdots \\ 0 \\ \mathcal{H}(u^{(L+1)}) \\ 0 \\ \vdots \\ 0
	\end{bmatrix} \alpha \\
 + \begin{bmatrix}
	\mathcal{H}(u) \\ \vdots \\ \mathcal{H}(u^{(L-1)}) \\ \mathcal{H}(u^{(L)}) \\ \mathcal{H}(y) \\ \vdots \\ \mathcal{H}(y^{(L-1)})
\end{bmatrix}^\dagger \begin{bmatrix}
		0 \\ \vdots \\ 0 \\ \bar{u}^{(L+1)} \\ 0 \\ \vdots \\ 0
	\end{bmatrix}\; .
\end{multline}
provides also a solution of \eqref{eq:difeq2}. This is an \emph{explicit} system of differential equations of the form $\dot \alpha(t) = \mathcal{A}(t) \alpha(t) + \nu(t)$ that corresponds to a linear time-varying dynamical system and, thus, can be solved using standard  software packages.

The full row rank requirement for the data matrix
\begin{equation} \label{eq:iodatamat}
	\begin{bmatrix}
		\mathcal{H}(u) \\ \vdots \\ \mathcal{H}(u^{(L-1)}) \\ \mathcal{H}(u^{(L)}) \\ \mathcal{H}(y) \\ \vdots \\ \mathcal{H}(y^{(L-1)})
	\end{bmatrix}
\end{equation}
on the left-hand side of \eqref{eq:difeq2} is only needed to conveniently solve \eqref{eq:simulation} using well-known methods, although any solution obtained directly from \eqref{eq:difeq2} is equally valid. Insights about sufficient conditions to guarantee that \eqref{eq:iodatamat} has full row rank can be obtained from  known results in the discrete-time case. For example, it was shown in \cite{AlsLopMul23} that a discrete-time data matrix, analogous to \eqref{eq:iodatamat}, can have full row rank only if $L=\ell ag(\B)$ and $p\, \ell ag(\B) = n$, where $p$ is the number of outputs and $n$ is the dimension of a minimal state of the system. Also in \cite{AlsLopMul23}, a method to obtain full row rank matrices when $p\, \ell ag(\B) > n$ was developed. This method relies on constructing non-minimal system states using input-output information. Pursuing these ideas for the continuous-time case is an interesting subject of future research.

\begin{remark}\rm 
    An important practical aspect of the described procedure is that performing these operations on digital computers requires the use of samples of data, instead of continuous-time signals. This allows for the numerical solution of \eqref{eq:difeq2} or \eqref{eq:simulation}. Moreover, even when using trajectory samples, our continuous-time system representation has computational complexity advantages with respect to using the discrete-time Willems lemma in \cite{WRM}. We refer the reader to \cite[Section~IV-C]{LM} for the details about these issues. 
\end{remark}

\section{Conclusions}\label{sec:concl}
We presented a novel version of Willems’ fundamental lemma for continuous-time systems and provided sufficient and necessary conditions for the data-based representation of system trajectories. The proposed approach overcomes the drawbacks of the existing methods in \cite{LM} and \cite{REtAl23b}, since it fully describes exact system trajectories using only external variables. We also showed how to use this system description to solve the continuous-time data-driven simulation problem. 

Future research directions include investigating the computation of non-minimal state variables for continuous-time systems and their application to data-based system representation.

\end{document}